\documentclass[aps,prd,preprint,floats,superscriptaddress,nofootinbib]{revtex4}
\usepackage[utf8]{inputenc}
\usepackage{amsmath,amssymb,bbm}
\usepackage{dsfont}
\usepackage{graphicx}
\usepackage{color,slashed}
\pdfoutput=1

\begin{document}
\title{\bf Structure of leptonic Yukawa couplings in the Zee model}

\author{Chuan-Hung Chen}
\email[E-mail: ]{physchen@mail.ncku.edu.tw}
\affiliation{Department of Physics, National Cheng-Kung University, Tainan 70101, Taiwan}

\date{\today}

%%%%%%%%%%%%%%%%%%%%%%%%%%%%%%%%%%%%%%%%%%%%%%%%%%
\begin{abstract}

The radiative neutrino mass matrix $m^\nu$ in the Zee model depends on leptonic Yukawa couplings $F$ to a singlet scalar and $Y^\ell$ to a new Higgs doublet. Leveraging the skew-symmetric structure of $F$, we derive a unique identity linking $F$ and $m^\nu$ that is explicitly independent of $Y^\ell$. This relation implies that five entries of $Y^\ell$ can, in principle, be determined directly from $m^\nu$ and $F$, while the remaining four can be selected based on phenomenological assumptions. As an illustration, we apply this framework to the two-zero texture $B2$, highlighting its enhancement of the muon $g-2$.

\end{abstract}
%%%%%%%%%%%%%%%%%%%%%%%%%%%%%%%%%%%%%%%%%%%%%%%%%%

\maketitle

%%%%%%%%%%%%%%%%%%%%%%%%%%%%%%%%%%%%%%%%%%%%%%%%%%
\section{Introduction} \label{sec:intro}
%%%%%%%%%%%%%%%%%%%%%%%%%%%%%%%%%%%%%%%%%%%%%%%%%%

The Majorana neutrino mass matrix can generally be expressed through the master formula~\cite{Cordero-Carrion:2018xre,Cordero-Carrion:2019qtu}:
\begin{equation}
m^\nu = \kappa (F\, M Y^\ell + Y^{\ell T}\, M^T F^T)\,, \label{eq:master}
\end{equation}
 where $F$ and $Y^\ell$ are Yukawa coupling matrices, $M$ is the fermionic mass matrix of the intermediate states, and $\kappa$ denotes model-dependent prefactors for tree-level, one-loop, or two-loop realizations.  
 
The Zee model, uniquely among neutrino mass models, features a skew-symmetric $F$ with three independent entries, corresponding to couplings of an $SU(2)_L$ singlet charged scalar to doublet leptons~\cite{Zee:1980ai}. This structure requires $Y^\ell$, which encodes couplings of the new scalars to leptons, to include flavor-changing neutral currents (FCNCs); otherwise, the vanishing of the diagonal elements of $m^\nu$ would conflict with neutrino oscillation data.

FCNCs in the lepton sector are tightly constrained by charged lepton-flavor violation (cLFV) processes, including tree-level muonium-antimuonium oscillations and trilepton decays $\ell_i \to \ell_j \ell_k \ell_l$, as well as loop-induced channels such as radiative decays $\ell_i \to \ell_j \gamma$ and $\mu-e$ conversion in nuclei. Interestingly, a value of $Y^\ell_{22}$ of order $0.1$, consistent with cLFV bounds, can significantly enhance the muon anomalous magnetic dipole moment ($g-2$)~\cite {Chowdhury:2022moc, Heeck:2023iqc}.  To accommodate neutrino oscillation data, cLFV bounds, and/or the muon  $g-2$, specific textures for $F$ and $Y^\ell$ are often assumed~\cite{Chowdhury:2022moc, Heeck:2023iqc, Herrero-Garcia:2017xdu, Primulando:2022vip}. 

An alternative parametrization, defining  $M Y^\ell = Z + Q$ with $F Q = Q^T F$, yields $m^\nu = F  Z - Z^T F$. Here,  the five-entry $Z$ is fixed by $F$ and $m^\nu$, while $Q$, irrelevant to $m^\nu$, introduces four parameters $q_i$~\cite{Machado:2017flo}, constructed from $F$ and $Y^\ell$. This parametrization is subsequently adopted in phenomenological studies~\cite{Heeck:2023iqc}.  While texture-based and $Z\textendash Q$ parametrizations aid numerical fits, they obscure the underlying relationships among $F$, $Y^\ell$, $m^\nu$, and cLFV constraints, with $Y^\ell$ appearing only indirectly through the $q_i$ in the latter.

To elucidate the underlying structure of $m^\nu$, $F$, and $Y^\ell$ without additional assumptions, we analyze Eq.~(\ref{eq:master}) in detail.  Two key features emerge: (i) the skew-symmetric $F$ defines a pseudovector $u$ satisfying $F\cdot u=0$, which leads to $u^T m^\nu u = 0$~\cite{Felkl:2021qdn}, establishing a relation between $F$ and $m^\nu$ that is independent of $Y^\ell$; (ii) only five of the nine entries in $Y^\ell$ can be determined by $m^\nu$ and $F$,  leaving four undetermined. Accordingly, we derive complete expressions for $Y^\ell$ in terms of $m^\nu$, $m_\ell$, and $F$, and identify viable choices for the undetermined entries under constraints such as $m_e \ll m_{\mu,\tau}$, muonium–antimuonium oscillations, $\mu \to 3e$, and $\mu \to e \gamma$.  We demonstrate that setting specific $Y^\ell$ entries to zero can evade cLFV constraints. As an illustration, the framework is applied to the two-zero texture $B2$.

The paper is organized as follows. Sec.~\ref{sec:couplings} briefly summarizes the relevant interactions in the Zee model. In Sec.~\ref{sec:correlation}, we derive the unique correlation between the skew-symmetric Yukawa matrix $F$ and $m^\nu$, and present compact relationships among new scalar couplings, $m^\nu$, and the charged-lepton masses. Guided by phenomenological considerations, we identify candidate entries of $Y^\ell$ that are insensitive to $m^\nu$. Their vanishing simplifies the numerical analysis. Sec.~\ref {sec:B2} applies the framework to realize the two-zero texture $B2$ and presents the associated results. A summary of the study is given in Sec.~\ref{sec:summary}. 

%%%%%%%%%%%%%%%%%%%%%%%%%%%%%%%%%%%%%%%%%%%%%%%%%%
\section{Leptonic Yukawa couplings and radiative neutrino mass matrix} \label{sec:couplings}
%%%%%%%%%%%%%%%%%%%%%%%%%%%%%%%%%%%%%%%%%%%%%%%%%%

In this section, we briefly review the relevant interactions in the Zee model. To radiatively generate Majorana neutrino masses without introducing additional exotic fermions,  the standard model (SM) is extended to include a second Higgs doublet, $H_2$, and a charged scalar singlet, $\chi^+$.  Lepton number is explicitly violated by the trilinear term $H^T_1 i \tau_2  H_2 \chi^-$ in the scalar potential, where $\tau_2$ is the second Pauli matrix.  Based on the $SU(2)_L\times U(1)_Y$ gauge symmetry, the relevant Yukawa couplings and the trilinear scalar interaction can be written as:
\begin{equation}
-{\cal L} \supset \bar L \left( y^\ell_1 H_1 + y^\ell_2 H_2 \right) \ell_R + L^T C i\tau_2 F L \chi^+ + \mu H^T_1  i\tau_2 H_2 \chi^- + H.c.\,, \label{eq:ints}
\end{equation}
where $y^\ell_{1,2}$ are leptonic Yukawa coupling matrices, $C$ is the charge-conjugation operator, and $F$ is a skew-symmetric Yukawa matrix. In the Zee model, $y^\ell_1$ and $y^\ell_2$ must not be aligned in order to fit the observed neutrino oscillation data. Since quark couplings play no role in this context, we do not discuss them further.

The Higgs flavor basis $H_1$ and $H_2$ is used in Eq.~(\ref{eq:ints}), where $H_1$ and $H_2$ are defined such that the neutral component of $H_1$ acquires a vacuum expectation value (VEV) responsible for electroweak symmetry breaking, while $H_2$ has a vanishing VEV~\cite{Davidson:2005cw}. Their components in the $SU(2)_L$ doublets can be expressed as $H^T_1=(G^+, (v+H^0_1+i G^0)/\sqrt{2})$ and $H^T_2=(H^+, (H^0_2+iA^0)/\sqrt{2})$. Here, $G^+$ and $G^0$ denote the Goldstone bosons, $H^0_{1,2}$ are CP-even neutral scalars, and $A^0$ is a CP-odd scalar. If CP-even and CP-odd scalars do not mix, $A^0$ corresponds directly to a mass eigenstate. The CP-even scalars $H^0_1$ and $H^0_2$, as well as the charged scalars $H^+$ and $\chi^+$, can mix. Their relations with the physical mass eigenstates can be parametrized as: 
\begin{equation}
\left(  \begin{array}{c}
   H^0_1 \\ 
    H^0_2\\ 
  \end{array} \right) = \left(  \begin{array}{cc}
   c_{\beta-\alpha}
 & s_{\beta-\alpha} \\ 
  -s_{\beta-\alpha} &  c_{\beta-\alpha}\\ 
  \end{array} \right)  \left(  \begin{array}{c}
 H \\ 
    h\\ 
  \end{array} \right)\,, ~ ~
  \left(  \begin{array}{c}
   H^+ \\ 
    \chi^+\\ 
  \end{array} \right) = \left(  \begin{array}{cc}
   c_\theta
 & - s_\theta \\ 
  s_\theta &  c_\theta\\ 
  \end{array} \right)  \left(  \begin{array}{c}
 H \\ 
    h\\ 
  \end{array} \right)\,, \label{eq:mix}
\end{equation}
with $c_{\beta-\alpha}=\cos(\beta-\alpha)$, $s_{\beta-\alpha}=\sin(\beta-\alpha)$, $c_\theta=\cos\theta$, and $s_\theta=\sin\theta$. Note that the mixing between $H^+$ and $\chi^+$ arises from the trilinear $\mu$ term in Eq.~(\ref{eq:ints}).

If the charged lepton mass matrix is diagonalized through $m^{\ell}_{\rm dia} = v/\sqrt{2}(V^{L\dagger}_\ell y^\ell_1 V^R_\ell)$, the Yukawa couplings of $H^0_{1,2}$, $H^-$, and $\chi^+$ to physical charged leptons can be written as:
 \begin{align}
 \begin{split}
 -{\cal L}^\ell_Y & = \bar\ell_R m^\ell_{\rm dia} \ell_L \left( 1+ \frac{H^0_1}{v}\right) + \bar\ell_R Y^\ell \ell_L (H^0_2  
 + i A^0 ) \\
 & +\sqrt{2}  \bar\ell_R Y^\ell \nu_L H^- + 2 \nu^T_L  C F \ell_L \chi^+ + H.c.\,,  \label{eq:Yukawa}
 \end{split}
 \end{align}
with $Y^\ell= ( V^{\ell \dagger}_L y^\ell_2 V^\ell_R/\sqrt{2})^\dagger$. As neutrino masses arise radiatively, the neutrinos appearing in the tree-level interactions of Eq.~(\ref{eq:Yukawa}) are effectively massless. 
Based on the interactions in Eq.~(\ref{eq:Yukawa}) and the charged scalar mixing angle $\theta$ defined in Eq.~(\ref{eq:mix}), the loop-induced Majorana neutrino mass matrix mediated by charged scalars can be expressed as:
%is radiatively generated via the one-loop Feynman diagram shown in Fig.~\ref{fig:zee_loop}.  The resulting neutrino mass matrix takes the %form:
\begin{equation}
m^\nu = \kappa (F m_\ell Y^\ell + Y^{\ell T}  m_\ell F^T)\,, \label{eq:mnu}
\end{equation}
where $m_\ell={\rm diag}(m_e ,m_\mu, m_\tau)$ is the diagonal charged-lepton mass matrix, and $\kappa$ is the loop factor, given by
 \begin{equation}
 \kappa= \frac{\sqrt{2} s_{2\theta}}{16\pi^2} \ln\left( \frac{m^2_{H^+_1}}{m^2_{H^+_2}}\right)\,.
 \end{equation}
Here, $s_{2\theta}=\sin2\theta$, and $m_{H^+_1}$ and $m_{H^+_2}$ denote the physical masses of $H^+$ and $\chi^+$, respectively.

%\begin{figure}[phtb]
%\begin{center}
%\includegraphics[scale=1.2]{zee_loop.pdf}
%\caption{ Feynman diagram illustrating the one-loop radiative generation of Majorana neutrino masses in the Zee model.}
%\label{fig:zee_loop}
%\end{center}
%\end{figure}

Introducing the unitary transformation $\nu_L \to U^\dagger \nu_L$, the Majorana neutrino mass matrix in Eq.~(\ref{eq:mnu}) can be diagonalized as $m^\nu_{\rm dia} = U^T m^\nu U$. The unitary matrix $U$ then appears in the weak charged-current interaction as $\bar\ell_L \gamma^\mu U \nu_L W^-_\mu$. After removing the unphysical phases of $U$, $U$ contains six independent parameters: three mixing angles, one Dirac CP-violating phase, and two Majorana phases. The standard parametrization of the mixing matrix $U$ is given as~\cite{ParticleDataGroup:2024cfk}:
   \begin{equation}
   U=
\left(
\begin{array}{ccc}
1 & 0 & 0 \\
 0 & c_{23} & s_{23} \\
 0 & -s_{23} & c_{23} \\
\end{array}
\right)\cdot\left(
\begin{array}{ccc}
 c_ {13} & 0 & s_{13} e^{-i\delta_{CP}} \\
 0 & 1 & 0 \\
 -s_{13} e^{-i\delta_{CP}}& 0 & c_{13} \\
\end{array}
\right)\cdot \left(
\begin{array}{ccc}
 c_{12} & s_{12} & 0 \\
 -s_{12} &c_{12} & 0 \\
 0 & 0 & 1 \\
\end{array}
\right)\cdot \left(
\begin{array}{ccc}
 e^{i\eta_1/2} & 0 & 0 \\
 0& e^{i\eta_2/2}& 0 \\
 0 & 0 & 1 \\
\end{array}
\right)~, \nonumber
   \end{equation}
   where $c_{ij}=\cos\theta_{ij}$, $s_{ij}=\sin\theta_{ij}$, $\delta_{CP}$ is the Dirac CP phase, and $\eta_{1,2}$ are the Majorana phases. 
 The mixing angles $\theta_{ij}$, the Dirac phase $\delta_{CP}$, and the mass-squared differences $m^2_{21}$ and $m^2_{3\ell}$  can be extracted from global fits to neutrino oscillation data~\cite{Esteban:2024eli}, where $\ell=1$ and $2$ are for normal ordering (NO) and inverted ordering (IO), respectively. However, since oscillation experiments are insensitive to the absolute neutrino mass scale and the Majorana phases, complementary probes are necessary. These include cosmological observations~\cite{Planck:2018vyg, DiValentino:2019dzu},  effective electron-neutrino mass observed by the beta-decay endpoint experiments~\cite{KATRIN:2024cdt}, and neutrinoless double beta decay ($0\nu\beta\beta$)~\cite{KamLAND-Zen:2022tow}.  Furthermore, assuming specific structures for the neutrino mass matrix $m^\nu$ with fewer matrix elements, such as the two-zero texture matrices~\cite{Frampton:2002yf, Xing:2002ta, Treesukrat:2025dhd}, can provide deeper insights into the nature of neutrinos if the proposed textures in $m^\nu$ yield a good global fit to experimental data.

%%%%%%%%%%%%%%%%%%%%%%%%%%%%%%%%%%%%%%%%%%%%%%%%%%
\section{Element-wise correlations among  $Y^\ell$, $F$, and $m^\nu$} \label{sec:correlation}
%%%%%%%%%%%%%%%%%%%%%%%%%%%%%%%%%%%%%%%%%%%%%%%%%%

The leptonic Yukawa couplings $Y^\ell$ and $F=[f_{ij}]$ are free parameters. 
Scanning the parameter space yields eigenvalues and eigenvectors of $m^\nu$, which can then be compared with neutrino oscillation data and cLFV constraints to determine the allowed parameter space. Since $Y^\ell$ contains more parameters than oscillation data can determine, it is convenient to work with a reduced parameter set for studying lepton-flavor phenomenology. This requires clarifying the role of $Y^\ell$ in shaping $m^\nu$ and cLFV processes within the Zee model before imposing further assumptions.

From the symmetry property of $m^\nu$ and $F$, we derive analytic formulations for $Y^\ell_{ij}$ in terms of $f_{ij}$, $m_\ell$, and $m^\nu_{ij}$, and identify which of its entries are insensitive to neutrino oscillation parameters. These results provide a systematic basis for applications to lepton phenomenology, especially when supplemented with suitable assumptions or approximations.

\subsection{The unique relation between $F$ and $m^\nu$}

Before presenting the relationships of $Y^\ell_{ij}$ in terms of $m^\nu_{ij}$, $m_\ell$, and $f_{ij}$, we first highlight a distinctive relation between $F$ and $m^\nu$ in the Zee model. From the skew-symmetric matrix $F$, a pseudovector can be defined as $u_i = \epsilon_{ijk} f_{jk}/2$, where $\epsilon_{ijk}$ is the three-dimensional Levi-Civita symbol. This construction ensures  $F\cdot u=0$.  Multiplying the neutrino mass matrix in Eq.~(\ref{eq:mnu}) on both sides by the column vector $u$ yields:
 \begin{equation}
  u^T m^\nu u = - \kappa\, u^T (F^T\; m_\ell\; Y^\ell + Y^{\ell T} \; m_\ell\; F) u= 0\,. \label{eq:umu_1}
 \end{equation}
An analogous application of the identity can be found in models with a singlet charged scalar~\cite{Felkl:2021qdn}. This identity leads to the following constraint:
 \begin{equation}
 f^2_{23} m^\nu_{11} + f^2_{13} m^\nu_{22}+ f^2_{12} m^\nu_{33} - 2 f_{13} f_{23} m^\nu_{12} -2 f_{12} f_{32} m^\nu_{13} -2  f_{21} f_{31} m^\nu_{23}=0\,,  \label{eq:umu_2}
 \end{equation}
which does not involve $Y^\ell$ explicitly.  We stress that Eq.~(\ref{eq:umu_1}) or Eq.~(\ref{eq:umu_2}) holds for any form of $Y^\ell$ without assuming a specific texture, although it arises from a particular pattern discussed in Ref.~\cite{Machado:2017flo}. 
This relation fixes one element of $m^\nu$ once the other five are specified by $Y^\ell$ and $F$.
% rather than a condition that determines any of the three elements of $f_{ij}$.

\subsection{Determination of $Y^\ell_{ij}$ from $f_{ij}$ and $m^\nu_{ij}$}

Disregarding the constraint in Eq.~(\ref{eq:umu_1}), the five independent elements of $m^\nu$ constrain the nine components of $Y^\ell$, leaving four entries undetermined. It is of interest to identify which four entries of $Y^\ell$ are insensitive to the neutrino mass matrix.
A natural choice, given the hierarchy $m_e \ll m_\mu < m_\tau$, is to assign these undetermined entries to terms multiplied by $m_e$. Since $m_e/m_{\mu(\tau)} \ll 1$, such terms become negligible and thus weakly constrained by neutrino data~\cite{Herrero-Garcia:2017xdu}. Guided by this observation, we aim to identify the four possible entries of $Y^\ell$.

Accordingly, from Eq.~(\ref{eq:mnu}), five complete relations among $Y^\ell_{ij}$, $m^\nu_{ij}$, $m_\ell$, and $f_{ij}$ are derived as:
\begin{subequations}
\begin{align}
2f_{13} m_\tau Y^\ell_{31} & =\kappa^{-1} m^\nu_{11} - 2 f_{12} m_\mu Y^\ell_{21}\,,  \label{eq:Y_F_m_a}\\
2 f_{23} m_\tau Y^\ell_{32}  & = \kappa^{-1} m^\nu_{22} + 2 f_{12} m_e Y^\ell_{12}\,,  \label{eq:Y_F_m_b}\\
2 f_{23} m_\mu Y^\ell_{23}  & = -\kappa^{-1} m^\nu_{33} - 2 f_{13} m_e Y^\ell_{13}\,,  \label{eq:Y_F_m_c}\\
2 f_{13} f_{23} ( m_\tau Y^\ell_{33}-m_e Y^\ell_{11}) & = \kappa^{-1} (f_{12} m^\nu_{33} +2 f_{23} m^\nu_{13})  \nonumber \\
& + 2 \left(f^2_{23} m_\mu Y^\ell_{21} + f_{12} f_{13} m_e Y^\ell_{13} \right) \,, \label{eq:Y_F_m_d}\\
f_{23} ( m_\tau Y^\ell_{33}-m_{\mu} Y^\ell_{22}) & = \kappa^{-1}  m^\nu_{23} + m_e (f_{13} Y^\ell_{12} + f_{12} Y^\ell_{13})\,. \label{eq:Y_F_m_e}
\end{align}
\label{eq:Y_F_m}
\end{subequations}
It is evident that $Y^\ell_{11}$, $Y^\ell_{12}$, and $Y^\ell_{13}$, which appear alongside $m_e$, could be insensitive to neutrino oscillation parameters, unless the couplings associated with $m_\mu$ and $m_\tau$ vanish or are sufficiently small. As shown in Eq.~(\ref{eq:Yukawa}), $Y^\ell$ contributes not only to charged currents but also to neutral currents. In particular, $Y^\ell_{12}$ and $Y^\ell_{21}$, mediated by CP-even and CP-odd scalars, can induce tree-level muonium-antimuonium oscillation~\cite{Heeck:2023iqc, Conlin:2020veq}. To avoid such tree-level cLFV processes, one may impose $Y^\ell_{12}\approx Y^\ell_{21}\approx 0$. Under the phenomenological assumptions $m_e\approx 0$ and $Y^\ell_{21}\approx 0$, Eq.~(\ref{eq:Y_F_m}) simplifies to
\begin{subequations}
\begin{align} 
2f_{13} m_\tau Y^\ell_{31} & =\kappa^{-1} m^\nu_{11} \,,  \\
2 f_{23} m_\tau Y^\ell_{32}  & = \kappa^{-1} m^\nu_{22} \,,  \\
2 f_{23} m_\mu Y^\ell_{23}  & = -\kappa^{-1} m^\nu_{33}\,,  \\
2 f_{13} f_{23}  m_\tau Y^\ell_{33}& = \kappa^{-1} (f_{12} m^\nu_{33} +2 f_{23} m^\nu_{13})  \,, \\
f_{23}  ( m_\tau Y^\ell_{33}-m_{\mu} Y^\ell_{22})& = \kappa^{-1}  m^\nu_{23}\,.
\end{align}
\label{eq:Y_F_m_approx}
\end{subequations}
Thus, under minimal assumptions, the five entries $Y^\ell_{22}$, $Y^\ell_{23}$, $Y^\ell_{31}$, $Y^\ell_{32}$, and $Y^\ell_{33}$ associated with $f_{ij}$ are fixed by five elements of $m^\nu$.  Equivalently, each $m^\nu_{ij}$ is determined by specific $Y^\ell_{ij}\textendash f_{ij}$ combinations. It is worth noting that $m^\nu_{12}$ is absent from Eq.~(\ref{eq:Y_F_m_approx}); however, it is implicitly encoded in Eq.~(\ref{eq:umu_2}). A complete treatment of $m^\nu$ therefore requires including  Eq.~(\ref{eq:umu_2}). 

According to the above analysis, the most suitable candidates for the four undetermined entries of $Y^\ell$ are $Y^\ell_{11}$, $Y^\ell_{12}$, $Y^\ell_{13}$, and $Y^\ell_{21}$. These couplings can, in principle, be constrained by the cLFV processes. On the other hand, if the goal is to minimize the number of free parameters to explore the phenomenological implications of the Zee model, one may consider the scenario with $Y^\ell_{11}=Y^\ell_{12}=Y^\ell_{13}=Y^\ell_{21}=0$. This assumption simplifies the analysis of flavor physics while leaving the neutrino sector unaffected. Under this scenario, the approximation $m_e=0$ is no longer required, and the simplified relations in Eq.~(\ref{eq:Y_F_m_approx}) remain valid. Additionally, the cLFV processes induced at the tree level, such as $\mu\to 3e$, $\tau\to (3e, \mu e^- e^+)$, and muonium-antimuonium oscillation,  can be forbidden.  Since $Y^\ell_{22}$ is not required to be small or vanish in this scenario, it can play a role in enhancing the muon $g-2$~\cite{Chowdhury:2022moc, Heeck:2023iqc}. 

Although nonvanishing values of $Y^\ell_{22}$, $Y^\ell_{31}$, $Y^\ell_{23}$, and $Y^\ell_{32}$ can induce the tree-level processes $\tau \to e \mu^- \mu^+$ and $\tau\to 3\mu$, couplings of order $|Y^\ell_{31}|$, $|Y^\ell_{23}|$, $|Y^\ell_{32}| \sim \mathcal{O}(10^{-4})$ with the scalar masses $\sim\mathcal{O}(100)$ GeV~\cite{Chowdhury:2022moc} can satisfy the projected sensitivity of $\mathcal{O}(10^{-10}\textendash 10^{-9})$ at Belle II~\cite{Belle-II:2022cgf}. However, the stringent constraint from the loop-induced decay $\mu\to e\gamma$ is governed by $m_\tau/m_\mu (Y^\ell_{23})^* (Y^\ell_{31})^*$~\cite{Herrero-Garcia:2017xdu}. To satisfy the current upper limit on the branching ratio (BR),  ${\cal B}(\mu\to e\gamma)<3.1\times 10^{-13}$, as measured by MEG/MEG II~\cite{MEGII:2023ltw}, one may simply impose either $Y^\ell_{31}=0$ or $Y^\ell_{23}=0$. According to Eq.~(\ref{eq:Y_F_m_approx}), these assumptions correspond to $m^\nu_{11}=0$ or $m^\nu_{33}=0$, respectively. This results in a one-zero texture for the neutrino mass matrix $m^\nu$, which remains consistent with current neutrino data~\cite{Priya:2025khf}. 

\section{Application to two-zero texture neutrino mass matrix} \label{sec:B2}

According to the recent analysis in Ref.~\cite{Treesukrat:2025dhd}, seven two-zero texture neutrino mass matrices remain compatible with current data within uncertainties. In addition to neutrino oscillation data, the study incorporates constraints from cosmology, the endpoint of the beta decay spectrum, and $0\nu\beta\beta$. These observables, together with their current upper limits, are given by
 $\sum_i m_i  =m_1+m_2+m_3 \leq 0.13\textendash 0.52$ eV~\cite{Planck:2018vyg, DiValentino:2019dzu}, 
 $m^{\rm eff}_{\nu_e} = \sqrt{\sum_i  |U_{ei}|^2 m^2_i } \leq 0.45$ eV~\cite{KATRIN:2024cdt}, 
$m_{ee}  =| \sum_i U^2_{ei} m_i | \leq  36\textendash 156$ meV~\cite{KamLAND-Zen:2022tow}, 
 where the quoted ranges for the upper limits on $\sum_i m_i$ and $m_{ee}$ reflect uncertainties in cosmological models and in theoretical calculations of nuclear matrix elements, respectively.  
 
 In the Zee model, a two-zero texture for the neutrino mass matrix can be realized by selecting a specific pattern of four nonzero entries in the Yukawa coupling matrix $Y^\ell$. As an illustration, we consider texture $B2$ with $m^\nu_{12}=m^\nu_{33}=0$. From earlier discussions, the texture $B2$ can be obtained by imposing $Y^\ell_{11}=Y^\ell_{12}=Y^\ell_{13}=Y^\ell_{21}=Y^\ell_{23}=0$. Accordingly, the neutrino mass matrix elements in  Eq.~(\ref{eq:Y_F_m_approx}) are expressed in terms of  the nonvanishing Yukawa couplings $Y^\ell_{22}$, $Y^\ell_{31}$, $Y^\ell_{32}$, and $Y^\ell_{33}$ as:
 \begin{align} 
 \begin{split}
m^\nu_{11} &= 2 \kappa f_{13} m_\tau Y^\ell_{31}  \,,  ~~  m^\nu_{22} = 2 \kappa f_{23} m_\tau Y^\ell_{32}  \,,  \\
m^\nu_{13} &= \kappa  f_{13} m_\tau Y_{33}  \,,~~ 
m^\nu_{23} = \kappa  f_{23}  ( m_\tau Y^\ell_{33}-m_{\mu} Y^\ell_{22})  \,.\\
 \end{split}
 \label{eq:B2}
\end{align}
From Eq.~(\ref{eq:umu_2}), the condition $m^\nu_{12}=0$ yields $f_{12} m_\mu Y^\ell_{22} + f_{13} m_\tau Y^\ell_{32}+ f_{23} m_\tau Y^\ell_{31}=0$. For numerical estimates, we set $\kappa=10^{-5}$, which encapsulates the effects of charged scalar masses and their mixing angle.  The parameters are constrained using the $3\sigma$ ranges from the global neutrino oscillation fit provided by NuFit 6.0~\cite{Esteban:2024eli}. To evaluate the muon $g-2$ deviation from the SM prediction, $\Delta a_\mu=a^{\rm exp}_{\mu} - a^{\rm SM}_{\mu}$, and the tree-level BRs for $\tau\to e\mu^-\mu^+$ and $\tau\to 3\mu$, we follow the results shown in Ref.~\cite{Herrero-Garcia:2017xdu} and use $m_{H^0_2}=150$ GeV and $m_{A^0}=200$ GeV. For illustration, benchmark values of $f_{ij}$ and $Y^\ell_{ij}$ for normal ordering (NO) and inverted ordering (IO) are shown in Table~\ref{tab:benchmark}, along with the resulting $\sum_i m_i$, $m^{\rm eff}_{\nu_e}$, $m_{ee}$, $\Delta a_\mu$, the lightest neutrino mass $m_{\rm light}$, and the BRs for the decays $\tau\to e\mu^- \mu^+$ and $\tau\to 3\mu$. The predicted $\Delta a_\mu$ is close to the $1\sigma$ upper value reported in the White Paper 2025 (WP25), $\Delta a_\mu =38(63)\times 10^{-11}$~\cite{Aliberti:2025beg}. Meanwhile, the BRs for $\tau\to e\mu^- \mu^+$ and $\tau\to 3\mu$ can be probed at the projected sensitivities of Belle II~\cite{Belle-II:2022cgf}.

 %%%%%%%%%%%%%%%%%%%%%%%%%%
\begin{table}[hptb]
\caption{Benchmark values of the parameters (Para) $f_{ij}$ and $Y^\ell_{ij}$ for NO and IO, satisfying global neutrino data fits within $3\sigma$. Corresponding observables (Obs), including $\sum_i m_i$, $m^{\rm eff}_{\nu_e}$, $m_{ee}$, the muon $g-2$, the lightest neutrino mass $m_{\rm light}$, and BRs for $\tau \to e \mu^- \mu^+$ and $\tau \to 3\mu$, are shown. }
\begin{tabular}{c|ccccccc} 
\hline \hline
Para ~~& ~~$f_{12}(10^{-4})$ ~~& ~~$f_{13}(10^{-2})$~~ & ~~$f_{23}(10^{-4})$~~& ~~$Y^\ell_{22}$ ~~&~~$Y^\ell_{31}(10^{-4})$~ & $Y^\ell_{32}(10^{-4})$~&~$Y^\ell_{33}(10^{-5})$  \\ \hline 
NO &   $3.20 -i\, 0.14$ & $0.24$ & $5.96$ & $-0.118$ & $7.69-i\, 1.64$ & $7.37$ & $7.98$
\\ 
IO & $2.25 - i\, 0.13 $ & $0.23$ & $5.17$  &  $-0.123$ & $-9.02-i\, 1.87$ & $9.08$ & $8.36$
\\ \hline\hline
Obs &  $\sum_i m_i$  & $m^{\rm eff}_{\nu_e}$ & $m_{ee}$ & $m_{\rm light}$   & $ \Delta a_\mu (10^{-9})$ & ${\cal B}(\tau\to e\mu^- \mu^+)$ ~& ~${\cal B}(\tau\to 3\mu)$ 
\\ \hline
NO &   $0.220$ eV & $0.068$ eV  & $0.068$ eV & $0.068$ eV & $1.01$ & $4.7\times 10^{-10}$ & $3.9 \times 10^{-10}$ \\
IO & $0.212$ eV & $0.076$ eV & $0.076$ eV & $0.059$ eV & $1.07$ & $8.5\times 10^{-10}$ & $6.2\times 10^{-10}$
\\ \hline \hline 
\end{tabular}
\label{tab:benchmark}
\end{table}
%%%%%%%%%%%%%%%%%%%%%%%%%

\section{Summary}\label{sec:summary}

We analyzed the relationships between the leptonic Yukawa matrices $F$ and $Y^\ell$ and the one-loop-induced neutrino mass matrix in the Zee model. We demonstrate a unique relation arising from the skew-symmetric nature of $F$ and the neutrino mass matrix, independent of $Y^\ell$. As a consequence, four entries of $Y^\ell$ cannot be determined from the neutrino mass matrix and $F$ alone. 

We identify $Y^\ell_{11}$, $Y^\ell_{12}$, $Y^\ell_{13}$, and $Y^\ell_{21}$ as viable candidates for these undetermined entries, with the first three multiplied by $m_e$. In the limit $m_e\to 0$, these couplings become irrelevant to neutrino oscillation parameters; however, phenomenological constraints, particularly from muonium-antimuonium oscillation, require $Y^\ell_{12, 21}$ to be suppressed. In principle, all other Yukawa couplings $Y^\ell_{ij}$ can be constrained by charged lepton-flavor violation processes. Imposing $Y^\ell_{11}=Y^\ell_{13}=0$ forbids the tree-level decays $\mu\to 3 e$ and $\tau\to (3e, \mu e^-e^+)$. Among the remaining five couplings, $Y^\ell_{22}$, $Y^\ell_{23}$, $Y^\ell_{31}$, $Y^\ell_{32}$, and $Y^\ell_{33}$, additional ones may be set to zero, depending on the phenomenological scenario under study. As an illustration, we apply the framework to the two-zero texture $B2$ neutrino mass matrix. When fitted to neutrino oscillation data, the predicted $\Delta a_\mu$ is close to the $1\sigma$ upper value of current measurements, and the BRs for $\tau\to (e \mu^- \mu^+, 3\mu)$ are within the reach of the projected sensitivities at Belle II.

%%%%%%%%%%%%%%%%%%%%%%%%%%%%%%%%%%%%%%%%%%%%%%%%%%
\section*{Acknowledgments}
This work was supported in part by the National Science and Technology Council, Taiwan, under Grant No. NSTC-114-2112-M-006-009.

\appendix

%%%%%%%%%%%%%%%%%%%%%%%%%%%%%%%%%%%%%%%%%%%%%%%%%%%%%%%%%%%%

\end{document}